\documentclass[10pt,twocolumn,oneside,letterpaper,conference]{IEEEtran}
\IEEEoverridecommandlockouts
\usepackage{cite}
\usepackage{amsmath,amssymb,amsfonts}
\usepackage{graphicx}
\usepackage{textcomp}
\usepackage{xcolor}
\usepackage{xcolor}

\usepackage{cite}
\usepackage{graphicx}
\usepackage{epstopdf}

\usepackage{amsmath}
\usepackage{times}
\usepackage{latexsym}
\usepackage{graphicx}
\usepackage{bm}
\usepackage{amssymb}
\usepackage{caption}
\usepackage{stfloats}
\usepackage{cases}
\usepackage{array}
\usepackage{setspace}
\usepackage{fancyhdr}
\usepackage{fancyhdr}
\usepackage{algorithm}
\usepackage{algorithmicx}
\usepackage{algpseudocode}
\usepackage{textcomp}
\usepackage{wasysym}
\usepackage{balance} %balance package 1
\usepackage{url}
\usepackage{xcolor}
\usepackage{subfigure}
\usepackage{stfloats}
\usepackage{enumerate}
\usepackage{array}
\usepackage{xcolor}
\usepackage{tikz}
\usepackage{booktabs}
\usepackage{colortbl}

\usepackage{amsmath}
\usepackage{amssymb}
\DeclareMathOperator*{\argmax}{argmax}

\def\BibTeX{{\rm B\kern-.05em{\sc i\kern-.025em b}\kern-.08em
	T\kern-.1667em\lower.7ex\hbox{E}\kern-.125emX}}

\begin{document}

\title{%Multi Agent Cooperation Reinforcement Learning for Cooperative Edge Caching in Fog Radio Access Network
	Cooperative Edge Caching via Multi Agent Reinforcement Learning in Fog Radio Access Networks
}

\author{
	\IEEEauthorblockN{Qi Chang$^{1}$,
		Yanxiang Jiang$^{1,2,*}$, Fu-Chun Zheng$^{1,2}$, Mehdi Bennis$^3$,
		and Xiaohu You$^1$}
	\IEEEauthorblockA{$^1$National Mobile Communications Research Laboratory,
		Southeast University, Nanjing 210096, China\\
		$^2$School of Electronic and Information Engineering, Harbin Institute of Technology, Shenzhen 518055, China\\
		$^3$Centre for Wireless Communications, University of Oulu, Oulu 90014, Finland\\
		%$^*$
		E-mail: $\{$220200705@seu.edu.cn, yxjiang@seu.edu.cn, fzheng@ieee.org, mehdi.bennis@oulu.fi, xhyu@seu.edu.cn$\}$
}}

\maketitle

\begin{abstract}
	%Fog radio access networks (F-RANs) can effectively alleviate network backhaul link congestion and reduce content transmission delay by migrating cloud services to edge networks.
	In this paper, the cooperative edge caching problem in fog radio access networks (F-RANs) is investigated. To minimize the content transmission delay, we formulate the cooperative caching optimization problem to find the globally optimal caching strategy.By considering the non-deterministic polynomial hard (NP-hard) property of  this problem, a Multi Agent Reinforcement Learning (MARL)-based cooperative caching scheme is proposed.Our proposed scheme applies double deep Q-network (DDQN) in every fog access point (F-AP), and introduces the communication process in multi-agent system. Every F-AP records the historical caching strategies of its associated F-APs as the observations of communication procedure.By exchanging the observations, F-APs can leverage the cooperation and make the globally optimal caching strategy.Simulation results show that the proposed MARL-based cooperative caching scheme has remarkable performance compared with the benchmark schemes in minimizing the content transmission delay.
\end{abstract}


\section{Introduction}

With the rapid advancement of wireless network technologies and the tremendous amount of data information, the global mobile data traffic generated by portable devices grows continuously in these years. Fog radio access network (F-RAN) has been proposed as a promising paradigm for improving spectral efficiency and optimizing legacy networks for mobile cellular communications systems\cite{AComprehensiveSurvey}\cite{SocialAware}. In F-RANs, edge caching can be regarded as a key component to relax the traffic burden at backhaul links by edge devices, e.g., fog access points (F-APs)\cite{Fog-computing-based}. Due to the finite cache capacity and communications resources of F-APs, the caching strategy should be designed comprehensively. In this regard, cooperative edge caching has become an efficient way to alleviate data traffic and decrease transmission delay.

There is a variety of research works focused on cooperative edge caching. 
%非学习方案 
In\cite{APigeonInspired}, an improved pigeon inspired optimization based cooperative edge caching scheme was proposed, which utilized Cauchy perturbation and self-adaptive factor to avoid premature convergence and achieve a better search performance. %an improved pigeon inspired optimization method was proposed to decompose the content transmission delay problem into two sub-problems solved separately.
In\cite{BrainStorm}, the authors proposed a brain storm optimization approach which utilized the penalty-based fitness function in individuals evaluation to meet the storage capacity constraint and the genetic algorithm in new individuals generation to meet the integer constraint, respectively. 
Specifically, with the maturation of reinforcement learning (RL), extensive works take RL into the optimization of cooperative edge caching.  
% 采用Qlearning
In\cite{Q-learning}, the authors deployed a distributed Q-learning based content replacement strategy, which created a Q-table to store the Q-value of every action. 
In\cite{LearningAutomata}, a learning automata based Q-learning algorithm for cooperative caching was proposed, which was invoked to obtain an optimal action selection at a random and stationary environment. 
% 采用DQN
In\cite{Dueling}, a delay-aware cache update policy was proposed in F-RANs with the dueling deep Q-network (DQN).  
In \cite{DistributedEdge}, the authors proposed a double DQN based distributed edge caching algorithm to find the optimal caching policy with content recommendation. 
In\cite{DeepReinforcement}, the cooperative caching problem was formulated by two potential recurrent neural networks, i.e., the echo state network and long short-term memory network, to determine which content to cache and where to cache. 
%采用MARL
By considering the leakage of sensitive users' data and additive waste of resources in training process, a cooperative caching method based on federated deep reinforcement learning framework was proposed  to find the optimal caching policy in\cite{CooperativeEdgeCaching}.
In\cite{Multi-Agent}, the authors extended Q-learning into multi-agent learning to solve the content transmission delay problem, which generally required complex computation for finding Nash-Q equilibrium.
Most of the aforementioned methods utilize RL to find the optimal caching strategy. %and a center controller is always deployed for scheduling the learning process. 
However, these RL-based methods generally neglect the influence of environment by other agents when a particular agent learns from the environment independently. 
%Although there are some methods, e.g.\cite{Multi-Agent} that have considered this problem, nash equilibrium in\cite{Multi-Agent} leads to high computational complexity. 

According to the above discussions, a cooperative edge caching scheme based on Multi Agent Reinforcement Learning (MARL) is proposed in F-RANs to find the globally optimal caching strategy.
Firstly, the cooperative edge caching optimization problem  is formulated to minimize the average transmission delay under the cache capacity and integer constraints.  
Then, double deep Q-network (DDQN) is utilized by each F-AP to learn how to coordinate their caching strategies in the multi-agent system.
Finally, every F-AP keeps its historical caching strategy as the observation of communication procedure.
Through the iterative communications among F-APs, the average transmission delay can be reduced and the optimization problem is tackled dynamically.

The rest of this paper is organized as follows. Section II introduces our system model and problem formulation. Section III describes the proposed MARL-based cooperative caching scheme. The simulation results are shown in Section IV. Finally, conclusions are drawn in Section V.

\begin{figure}[t]
	\centering 
	\vspace*{13pt}
	\includegraphics[height=5.5cm,width=7.5cm]{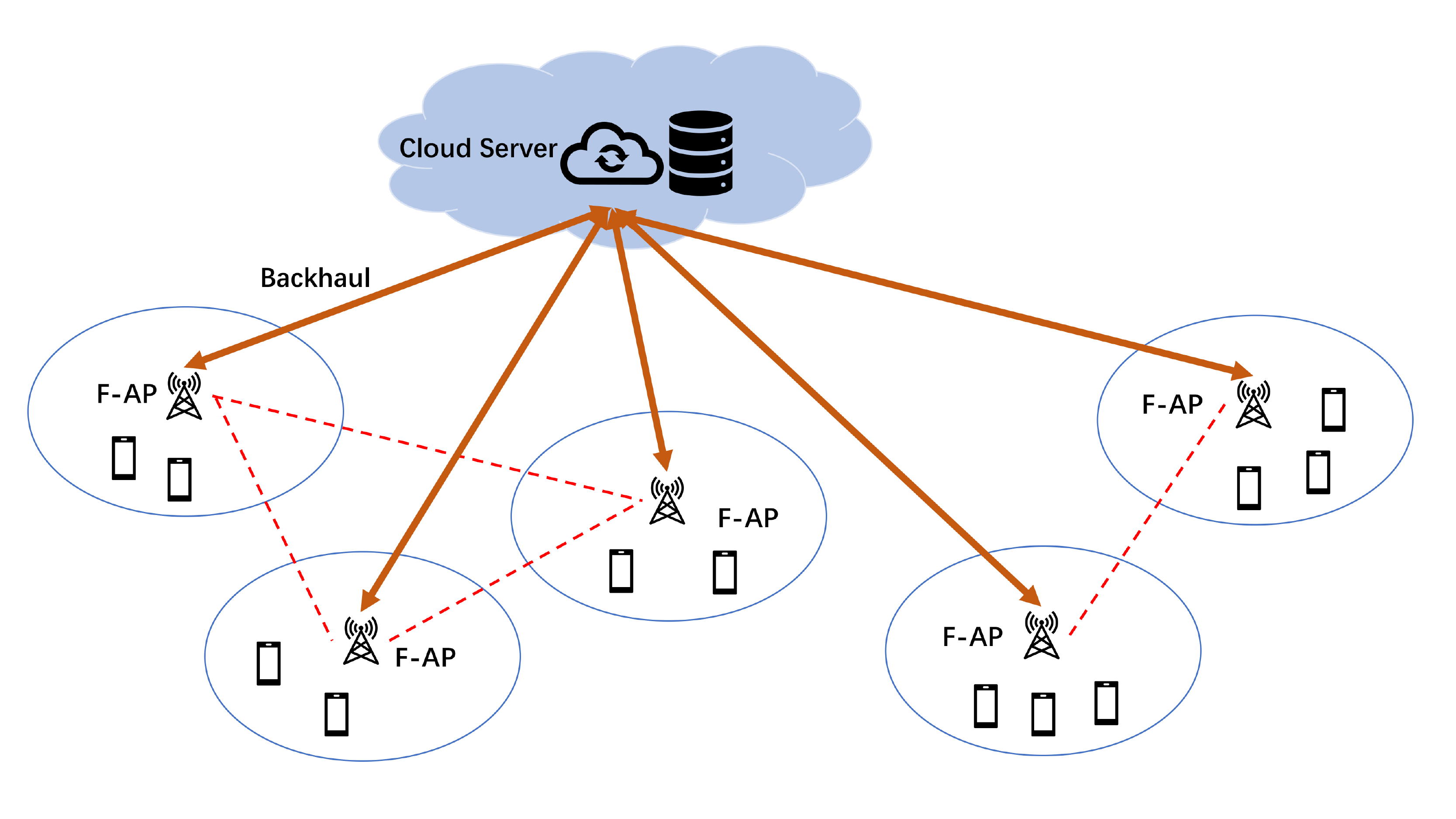}
	%\captionstyle{mystyle3}
	\caption{The cooperative caching scenario in F-RANs.}
	\label{scenario}
	%\vspace*{13pt}
	\vspace{-0.5cm}  %调整图片与上文的垂直距离
\end{figure}

\section{System Model and Problem Formulation}

\subsection{System Model}

The cooperative caching scenario in F-RANs is illustrated in Fig. 1, where a cloud server is connected with multiple F-APs via backhaul links and multiple users are under the serving region of each F-AP. The continuous time is divided into discrete time slots $\mathsf{\mathcal{T}}\text{=}\left\{1,2,...,t,...,T\right\}$. The set of F-APs is denoted by $\mathsf{\mathcal{N}}\text{=}\left\{1,2,...,n,...,N\right\}$ and the set of all the considered users is denoted by  $\mathsf{\mathcal{U}}\text{=}\left\{1,2,...,u,...,U\right\}$. The set of users in the serving region of F-AP ${n}$ is denoted by $ {\mathcal{U}_{n}^{t}}\text{=}\left\{1,2,...,u_n,...,U_n\right\}$. We assume that user ${u_n}$ is only served by F-AP ${n}$ during time slot ${t}$. 

Suppose that the library, denoted by $\mathsf{\mathcal{F}}\text{=}\left\{1,2,...,f,...,F\right\}$, is located at the cloud server far away from users, which can be accessed by F-APs via backhaul links. Furthermore, we assume that every file has the same size ${Q}$. 
The content popularity distribution in the serving region of F-AP ${n}$ is denoted by ${\mathcal{P}_{n}^{t}\text{=}\left\{{P}_{n,1}^{t},{P}_{n,2}^{t},...,{P}_{n,f}^{t},...,{P}_{n,F}^{t}\right\}}$.
Let ${{p}_{u,f}^{t}}$ denote the file preference of user $u$ for file $f$, which can be viewed as content popularity indicator and predicted via some learning procedure \cite{CachePlacement}.
We assume that the user’s file preference ${{p}_{u,f}^{t}}$ satisfies the Mandelbrot-Zipf distribution \cite{Citations} as follows:
\begin{equation}
	{p}_{u,f}^{t}\text{=}\dfrac{\phi_{u}^{t}\left({f}\right)^{-\tau_t}}{\sum\limits_{i=1}^{F} {{i}^{-\tau_t}}},\forall u\in \mathcal{U} ,
\end{equation}
where ${\phi_{u}^{t}\left(f\right) \in \mathcal{\Phi}_u^t = \left\{ \phi_{u}^{t}\left(1\right),\phi_{u}^{t}\left(2\right),...,\phi_{u}^{t}\left(f\right),...,\phi_{u}^{t}\left(F\right)\right\}}$, $ \mathcal{\Phi}_u^t$ is a random permutation of content library $\mathsf{\mathcal{F}}$ for user $u$ during time slot $t$, and ${\tau_t}$ is the time-varying skewness factor. The content popularity in F-AP $n$ generally depends on the file preference of its serving users $u_n \in \mathcal{U}_n$, and it can be calculated by:
\begin{equation}
	{{P}_{n,f}^{t}\text{=}\mathbb{E}_{u}[{\sum\limits_{{u}\in {\mathcal{U}_{n}}}{p}_{u,f}^{t}}]} ,
\end{equation}
where $\mathbb{E}\left[\cdot\right]$ denotes the operation of mathematical expectation.
We also assume that all F-APs have the same cache capacity ${S}$. %Denoted by ${S}$ the cache space is normalized, which means that each F-AP can cache at most ${S}$ files.
Let the binary variable ${{x}_{n,f}}$ indicate whether F-AP ${n}$ has cached file ${f}$. ${{x}_{n,f}\text{=}1}$ if file ${f}$ has been cached at F-AP ${n}$, and otherwise ${{x}_{n,f}\text{=}0}$. The caching variable ${{x}_{n,f}}$ should be determined collaboratively by all F-APs and the cooperative caching strategy, denoted by ${\boldsymbol{X}}=\left[{{x}_{n,f}}\right]_{N\times F}$, should be designed carefully to make file requests from all users respond quickly and accurately.

\subsection{Transmission Mode}
\setlength{\parskip}{0.1cm plus4mm minus3mm}
At the network edge, some F-APs can deliver the requested file via backhaul links\cite{AMeanField}. 
The connectivity among F-APs can be denoted by an ${N}\times{N}$ matrix ${\boldsymbol{Y}}$, where every binary element ${y}_{n,m}$ indicates whether F-AP ${n}$ can associate with F-AP ${m}$. 
${{y}_{n,m}\text{=}1}$ if F-AP ${n}$ can establish connection with F-AP ${m}$, and otherwise ${{y}_{n,m}\text{=}0}$. 
Therefore, the set of the associated F-APs for F-AP ${n}$ can be denoted by $\mathcal{N}_{n}\text{=}\left\{m|{\forall{m}}\in \mathcal{N}, {y}_{n,m}\text{=}1, {m}\ne {n} \right\}$.

When user ${u_n}$ requests file ${f}$, the serving F-AP $n$ checks its own caching strategy ${\left[{x}_{n,1},...,{x}_{n,f},...,{x}_{n,F}\right]}$ to decide how to transmit the requested file ${f}$ to user ${u_n}$. 
Some specific transmission modes are applied to deliver the file for the requesting user. In the following, we discuss the transmission delay with different transmission modes, when the requested file is cached in the serving F-AP, its associated F-APs or the cloud server.

\subsubsection{F-AP-to-User} If the requested file is cached in the serving F-AP, it can directly deliver the file to the requesting user. Let ${R}_{n,u,f}^{t}$ denote the delivery rate of file $f$ from F-AP $n$ to user $u_n$ during time slot $t$. Assume that efficient interference management schemes are applied and interference power is constrained by a fixed value ${{P}_{I}}$ \cite{CachePlacement}. Then, the file delivery rate in wireless transmission stage can be expressed as:
\begin{equation}
	\begin{aligned}
		{R}_{n,u,f}^{t}\text{=}{B}\log \left(1+{{|{h}_{n,u}^{t}|^2}{{l}_{n,u}^t}}\dfrac{{P_{n}}}{{{N}_{0}{{B}}}+{{P}_{I}}}\right) ,
		%&{R}_{i,a,b,f}^{t}\text{=}{{B}_{i}}\log \left(1+{{|{h}_{a,b}^{t}|^2}{{l}_{a,b}}}\dfrac{{P_{a}}}{{{N}_{0}{{B}_{i}}}+{{P}_{I}}}\right)\\
		%&{i}\text{=}{\left\{1,2,3\right\}},
		%(a,b)=\left\{
		%\begin{array}{rcl}
		%		&(n,{u\in\mathcal{U}_{n}})  & {i \text{=}1}\\
		%		&({m\in\mathcal{N}_{n}},n)  & {i \text{=}2}\\
		%		&(0,n)  & {i \text{=}3}\\
		%
		%		\end{array} \right.
	\end{aligned}
\end{equation}
where ${{B}}$ is the channel bandwidth, ${P}$ is the transmit power, ${{N}_{0}}$ is the power spectral density of noise, ${h}_{n,u}^{t}$ denotes the channel coefficient between F-AP ${n}$ and user ${u}$ during time slot ${t}$, and ${l}_{n,u}^t$ is the distance between F-AP ${n}$ and user ${u}$ during time slot ${t}$. 
Thus, the corresponding transmission delay can be defined as:
\begin{equation}
	{Z}_{1,n,u,f}^{t}\text{=}{Q}/{{R}_{n,u,f}^{t}} .
\end{equation}

\subsubsection{F-AP-to-F-AP} If the requested file is not cached in the serving F-AP, the requesting user can obtain the requested file from the associated F-APs that have cached the file. And the transmission process can be divided into two parts: the transmission delay from F-AP to the requesting user, i.e., ${Z}_{1,n,u,f}^{t}$ and the transmission delay between F-APs, i.e., ${Z}_{2,n,f}^{t}$. Then, we have:
\begin{equation}
	{Z}_{2,n,f}^{t}\text{=}{Q}\left(\sum\limits_{{m}\in\mathcal{N}_{n}}\dfrac{{x_{m,f}}}{{R}_{n,m,f}^{t}}\right) ,
\end{equation}
where ${R}_{n,m,f}^{t}$ is the transmit rate between F-APs. When there exist multiple associated F-APs that have stored the requested file, these associated F-APs can transmit the requested file cooperatively to improve the transmission performance\cite{BrainStorm}.  

\subsubsection{Cloud-Server-to-F-AP}If the requested file is cached neither in the serving F-AP nor in its associated F-APs, the requested file can only be fetched from the cloud server. And the transmission process can also be divided into two parts: the transmission delay from F-AP to the requesting user, i.e., ${Z}_{1,n,u,f}^{t}$ and the transmission delay from the cloud server to F-AP, i.e., ${Z}_{3,n,f}^{t}$. And the transmission delay in backhaul link can be defined as:
\begin{equation}
	{Z}_{3,n,f}^{t}\text{=}{C}/{{R}_{n,0,f}^{t}},
\end{equation}
where ${R}_{n,0,f}^{t}$ is the transmit rate from the cloud server to F-AP.

Based on the above discussions, the transmission delay for the requested file ${f}$ in three transmission modes can be expressed as:
\begin{equation}
	\begin{aligned}
		&{d}_{n,f}^{t}\left(\boldsymbol{X}\right)\text{=}{x}_{n,f}{{Z}_{1,n,u,f}^{t}}\\
		&+\left(1-{{x}_{n,f}}\right)\left(1-{\prod\limits_{{m}\in\mathcal{N}_{n}}{\left(1-{{x}_{m,f}}\right)}}\right)\left({{Z}_{1,n,u,f}^{t}}+{{Z}_{2,n,f}^{t}}\right)\\
		&+\left(1-{{x}_{n,f}}\right)\prod\limits_{{m}\in\mathcal{N}_{n}}{\left(1-{{x}_{m,f}}\right)}\left({{Z}_{1,n,u,f}^{t}}+{{Z}_{3,n,f}^{t}}\right).	
	\end{aligned}	
\end{equation}
Without loss of generality, ${{Z}_{1,n,u,f}^{t} < {Z}_{2,n,f}^{t} \ll {Z}_{3,n,f}^{t}}$ is assumed. If ${x}_{n,f} \text{=}1$, the requested file can be directly fetched from the serving F-AP. If ${x}_{n,f} \text{=}0$ and ${\prod\limits_{{m}\in\mathcal{N}_{n}}{\left(1-{{x}_{m,f}}\right)}}=0$, the requested file can be fetched from the associated F-APs. And if ${x}_{n,f} \text{=}0$ and ${\prod\limits_{{m}\in\mathcal{N}_{n}}{\left(1-{{x}_{m,f}}\right)}}=1$, the requested file can be fetched from the cloud server.
\subsection{Problem Formulation}
By considering time-varying channel state, diverse content preference of user and cooperation among F-APs, our work aims at finding the globally optimal caching strategy $\boldsymbol{X}^*$ to minimize the average transmission delay of the entire system. 
According to the transmission delay given by (7), the cooperative caching problem can be formulated as follows:
\begin{equation}
	\begin{aligned}
		\min_{{x}_{n,f}}\quad {\bar{D}\left(\boldsymbol{X}\right)}={\dfrac{1}{T}}\sum\limits_{t=1}^T\sum\limits_{f=1}^F\sum\limits_{n=1}^N {{{P}_{n,f}^{t}}\cdot{d}_{n,f}^{t}\left(\boldsymbol{X}\right)}\\
		{\rm s.t.}\quad  \left\{\begin{array}{lc}
			\sum\limits_{f=1}^F {{x}_{n,f}} \leq {S},  \forall n \in \mathcal{N},     &\left(8 \rm a \right) \\
			{{x}_{n,f}}\in \left\{0,1\right\},  \forall n \in \mathcal{N}, \forall f \in \mathcal{F},   & \left(8 \rm b \right) \\
		\end{array}\right.
	\end{aligned}	
\end{equation}
where the constraint (8a) implies that each F-AP is allowed to cache at most ${S}$ files, and the constraint (8b) implies that the caching strategy variable is binary.
%The objective of this paper is to find the optimal cooperative caching strategy $\boldsymbol{X}^*$ by minimizing the average transmission delay with cache capacity and integer constraints.

\section{Proposed MARL-based Cooperative Caching Scheme}

%In this section, we propose an MARL-based caching strategy to solve 
The optimization problem in (8) is a constrained integer programming problem and non-deterministic polynomial hard (NP-hard), which generally requires exponential computational complexity for traditional simple searching approaches to obtain the globally optimal solution \cite{BrainStorm}. 
%DDQN in RL can be applied as an effective approach to solve the problem.
%MARL has exhibited great potential for solving these problems because of its experience sharing.
To solve the problem with low computational complexity, we propose an MARL-based cooperative caching scheme.
We briefly introduce the DDQN in every F-AP to minimize the local transmission delay. However, individual training in the DDQN neglects the interaction among F-APs and cannot guarantee the minimum average transmission
delay of the entire system.
We then resort to MARL to build a communication procedure to leverage the cooperation among F-APs. By the joint learning of agents, the maximum global reward function is achieved and the average transmission delay of the entire system is minimized.  %, in which every F-AP learns from the communication procedure to maximize the global reward function, to find the globally optimal caching strategy in (8).

\subsection{Reinforcement Learning Framework}
%Among deep reinforcement learning algorithms, the deep Q-network (DQN) can effectively combine deep learning and reinforcement learning. The main reason is the DQN has three elements:
%\subsubsection{Objective function}Based on the Q-learning algorithm, an objective function learned by deep learning is constructed.
We model the local transmission process in single F-AP as a Markov Decision Process (MDP) with state space, action space and reward function. 
In detail, agent $n$ observes a state  ${\boldsymbol{s}_{n}^{t}}$ from the environment and executes an action ${a}_{n}^{t}$ during time slot $t$. Then, the environment feeds back a reward  ${r_n^t}\text{=}{r\left(\boldsymbol{s}_{n}^{t},{a}_{n}^{t} \right)}$ and the new state ${\boldsymbol{s}_{n}^{t+1}}$ to the agent. 
To employ the RL framework, the critical elements in MDP are identified as follows:
  
 %In detail, the agent will get a reward ${r_n^t}\text{=}{R\left(\boldsymbol{s}_{n}^{t},{a}_{n}^{t} \right)}$ from the environment after the selected action is executed.

\subsubsection{\textbf{State Space}}The state ${\boldsymbol{s}_{n}^{t} \in \mathcal{S}_n}$ indicates the cache status information of the $n$-th agent during time slot $t$ and the cache status can be denoted by ${\boldsymbol{s}_{n}^{t} \text{=}\left\{\boldsymbol{q}_{n}^{t},{f}_{n}^{t}\right\}}$. 
The former element $\boldsymbol{q}_{n}^{t}\text{=}\left\{{q}_{n,1}^{t},{q}_{n,2}^{t},...,{q}_{n,S}^{t}\right\}$ collects the indexes of cached files in agent $n$, which corresponds to the local caching strategy of F-AP $n$. The latter element ${f}_{n}^{t}\in \mathcal{F}$ is the requested file from the requesting user in the region of F-AP $n$. %Above information is assembled as a state of the agent.

\subsubsection{\textbf{Action Space}}The objective of an agent is to map the space of states to the space of actions. The action of agent $n$ is denoted by ${{a}_{n}^{t} \in \mathcal{A}_n}$. Let ${{a}_{n}^{t}\text{=} 0,1,...,S}$, where  ${{a}_{n}^{t}\text{=}s\left(s \ne 0\right)}$ means that the $s$-th cached content in F-AP $n$ will be replaced by the requested file ${f}_{n}^{t}$, and ${{a}_{n}^{t}\text{=}0}$ means that the requested file ${f}_{n}^{t}$ should not be cached. Then, the agent can update its own caching strategy according to the selected action.

\subsubsection{\textbf{Reward Function}} When agent ${n}$ selects an action ${{a}_{n}^{t}}$ under the state ${\boldsymbol{s}_{n}^{t}}$, a reward function ${r_n^t}$ is determined. The objective of RL is to obtain the minimum local transmission delay of F-AP $n$ and to achieve the maximum reward. Thus, the reward function is designed as follows:
%${r_n^t}\text{=} {{p}_{nf}}{{e}^{-{{\lambda }_{1}}{{d}_{n1}}}}$
\begin{equation}
	\begin{aligned}
		{r_n^t}\left(\boldsymbol{X}\right)\text{=}\sum\limits_{{f}\text{=}{1}}^{F}{{P}_{n,f}^{t}}{{e}^{-\lambda({d}_{n,f}^{t}\left(\boldsymbol{X}\right)-{{Z}_{1,n,u,f}^{t}})}},\\
		%{r_n^t}\left(\boldsymbol{X}\right)\text{=}{\rm exp}\left[\sum\limits_{{f}\text{=}{1}}^{F}{}{{-\lambda{P}_{n,f}^{t}({d}_{n,f}^{t}\left(\boldsymbol{X}\right)-{{Z}_{1,n,u,f}^{t}})}}\right],
	\end{aligned}
\end{equation}
where the exponential function is used to keep the reward function bigger than 0, and ${\lambda \, (0<\lambda\leq1)}$ guarantees that the reward function is normalized.

Besides, the optimal action-value function $Q_n^t(\boldsymbol{s}_{n}^{t},{a}_{n}^{t})$ in agent $n$ can be defined as follows:
\begin{equation}
	\begin{aligned}
		{Q_n^t(\boldsymbol{s}_{n}^{t},{a}_{n}^{t})}\leftarrow{Q_n^t(\boldsymbol{s}_{n}^{t},{a}_{n}^{t})}+\alpha [{r_n^{t+1}}\\+\gamma \max_{{a}_{n}^{t+1}}Q_n^t(\boldsymbol{s}_{n}^{t+1},{a}_{n}^{t+1})
		&-Q_n^t(\boldsymbol{s}_{n}^{t},{a}_{n}^{t})],
	\end{aligned}
\end{equation}
where $\alpha$ and $\gamma$ denote learning rate and reward decay respectively. %$Q_n^t(\boldsymbol{s}_{n}^{t},{a}_{n}^{t})$ is updated iteratively to obtain the optimal caching strategy $\boldsymbol{X^*}$. 
%The goal of agent is to find a caching strategy to achieve the maximum average reward.
%According (10), RL can find the locally optimal caching strategy in single F-AP and maximize the local reward. Generally, although MDP can be solved by dynamic programming algorithms, the curse of dimensionality occurs when the size of the optimal problem is huge. DDQN can be applied to improve the learning efficiency and maximize the reward function in (9). %a DDQN can be applied to solve the local caching strategy.
%Generally, although MDP can be solved by dynamic programming algorithms, the curse of dimensionality occurs when the size of the optimization problem is huge. Thus, DDQN can be applied to improve the learning efficiency and maximize the reward function of every F-AP.
%\subsection{Double Deep-Q-network}

\subsection{Double Deep Q-Network}
%Although the problem above can be solved by dynamic programming algorithms,  occurs when the size of the optimization problem is huge.  
RL techniques such as DQN and DDQN are applied as the effective approaches to tackle the curse of dimensionality and achieve the maximum reward. In addition, compared with DQN algorithm, DDQN can decouple the action selection from the calculation in (10) to prevent the overoptimistic value estimates\cite{DoubleQlearning}.
Correspondingly, DDQN based on RL is utilized to find the optimal strategy.
%is utilized for making the optimal strategy, which has two separate neural networks, a current Q-network and a target Q-network. 
In the architecture of DDQN, there are two separate neural networks, a current Q-network and a target Q-network. The current Q-network $Q_n^t(\boldsymbol{s}_{n}^{t+1},{a}_{n}^{t+1}|{\theta}_n)$ with the network parameter ${\theta}_n$ is utilized for approximating ${Q_n^t(\boldsymbol{s}_{n}^{t},{a}_{n}^{t})}$ in (10).
And the target Q-network ${\hat{Q}_n^t(\boldsymbol{s}_{n}^{t},{a}_{n}^{t}|\hat{\theta}_n)}$ with the network parameter $\hat{\theta}_n$ is utilized for computing the target Q-value. It can be expressed as follows:
\begin{equation}
	\begin{aligned}
		{\hat{Q}_n^t(\boldsymbol{s}_{n}^{t},{a}_{n}^{t}|\hat{\theta}_n)} \text{=}{r_n^{t+1}}+\gamma \hat{Q}_n^t(\boldsymbol{s}_{n}^{t}, a'|\hat{\theta}_n),
	\end{aligned}
\end{equation}
where $a'={\argmax_{{a}_{n}^{t+1}}Q_n^t(\boldsymbol{s}_{n}^{t+1},{a}_{n}^{t+1}|{\theta})}$ is an action chosen from the current Q-network to maintain the current Q-value under the state $\boldsymbol{s}_{n}^{t+1}$, and $\hat{\theta}_n $ is the weight of the $n$-th target Q-network.

Instead of updating the network parameters of the target Q-network iteratively, they are copied from the current Q-network at intervals, i.e., delayed update, which reduces the correlation between the target Q-value and the current Q-value. The loss function in the network is updated via a gradient descent approach as follows\cite{CooperativeEdgeCaching}:
\begin{equation}
	\begin{aligned}
		{L({\theta}_n)} \text{=}(\hat{Q}_n^t(\boldsymbol{s}_{n}^{t},{a}_{n}^{t}|\hat{\theta}_n)-Q_n^t(\boldsymbol{s}_{n}^{t},{a}_{n}^{t}|{\theta}_n))^2,
	\end{aligned}
\end{equation}
where the current Q-network parameters $\theta_n$ can be obtained according to (12), and the target Q-network parameter $\hat{\theta}_n$ will copy $\theta_n$ from the current Q-network $Q_n^t(\boldsymbol{s}_{n}^{t},{a}_{n}^{t}|{\theta}_n)$ every $\nu$ steps.
%\subsubsection{Experience Reply}The DDQN introduces an experience reply mechanism to store the experience sample data obtained by the interaction among agents and the environment at each time slot. 

%Each experience data is stored as $\left[\boldsymbol{s}_{n}^{t},{a}_{n}^{t},R^t,\boldsymbol{s}_{n}^{t+1},C_{-n,f}^t\right]$. It indicates that the learning agent $n$ reaches the next state $\boldsymbol{s}_{n}^{t+1}$ after executing the action ${a}_{n}^{t}$ under the current state $\boldsymbol{s}_{n}^{t}$ and the observation $C_{-n,f}^t$. Then, agent $n$ obtains the global reward $R^t$.

%Besides, DDQN uses the experience replay $\mathcal{D}$ to store the experience ${\boldsymbol{e}_n^t\text{=}\left[\boldsymbol{s}_{n}^{t},{a}_{n}^{t},{r}_{n}^{t},\boldsymbol{s}_{n}^{t+1}\right]}$, from which small batches of data are selected randomly to update the current Q-network. % every ${\nu}$ time slots. 

\subsection{Proposed MARL-based Cooperative Caching Scheme}
%解释MARL
In the above work, we have utilized the DDQN in single F-AP. 
%Nevertheless, conventional RL model ignores the influence of other agents' states, which means the relationship among agents is independent. 
%However, DDQN has neglected the interaction among agents. The maximum reward of every F-AP in (9) cannot guarantee the minimum average transmission delay of the entire system. 
In order to leverage the cooperation among F-APs, we extend DDQN to multi-agent system and introduce the communication procedure among F-APs, which is illustrated in Fig. 2.

\begin{figure}[t]
	\centering %\vspace*{135pt}
	\includegraphics[height=5cm,width=8.5cm]{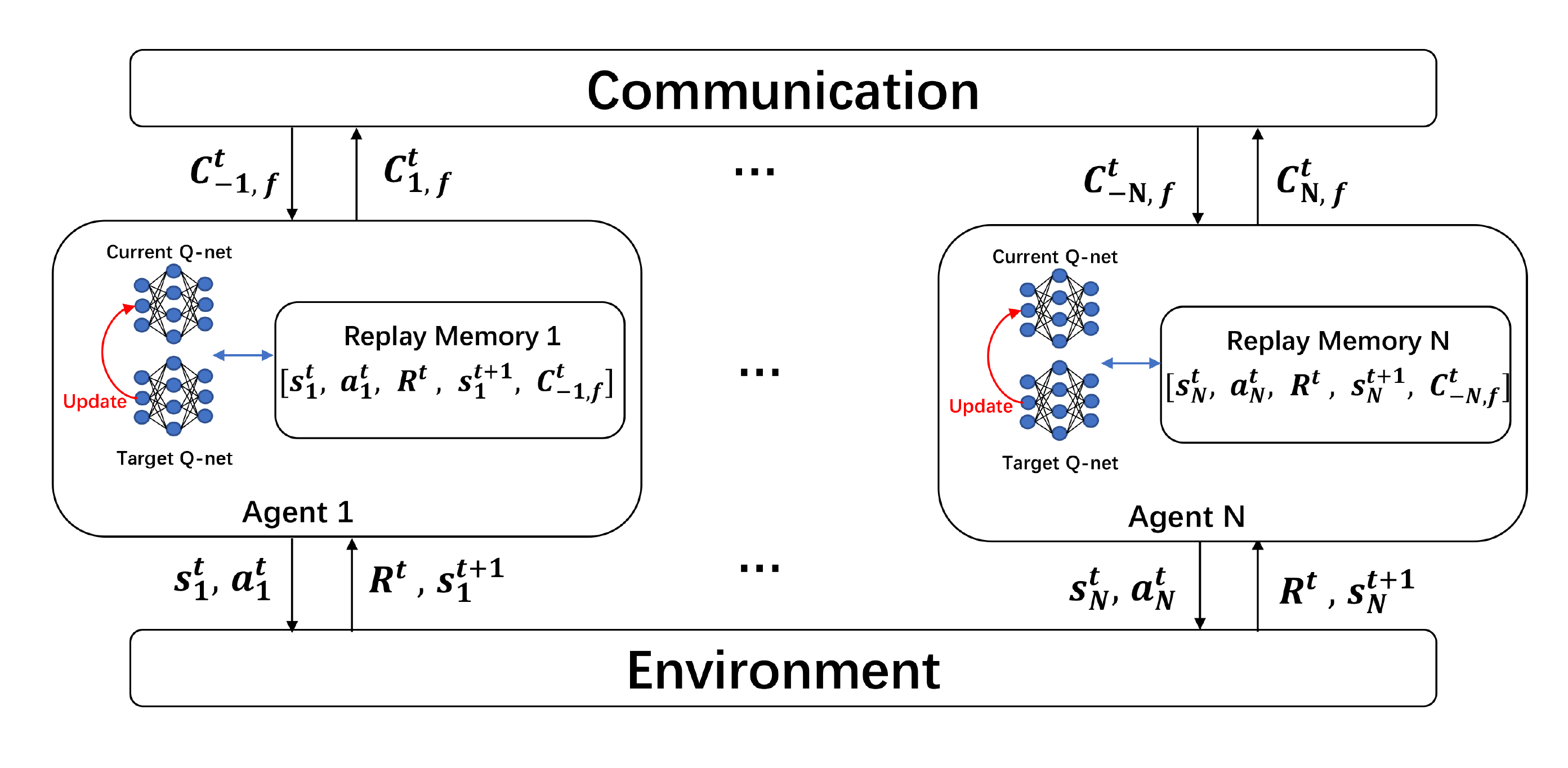}
	%\captionstyle{mystyle3}
	\caption{Schematic of the MARL framework.}
	\vspace{-0.5cm}
	%\label{scenario}
\end{figure}

The global caching strategy can be formulated as Stochastic Game (SG) \cite{Multi-Agent}. The SG model can be defined as ${\left\{N,{S_1\times ... \times S_N},{A_1\times ... \times A_N},R^t\right\}}$, where ${S_n}$ is the state space of the $n$-th agent, ${A_n}$ is the action space of the $n$-th agent, and $R^t$ is the global reward function. So the joint action space is ${\mathcal{A}={A_1\times ... \times A_N}}$ and the joint state space is ${\mathcal{S}={S_1\times ... \times S_N}}$. Since every agent's action has an impact both on the local reward as well as on the global reward, all agents are expected to work cooperatively to find the globally optimal strategy that maximizes the global reward. 
By considering the reward function ${r_n^t}\left(\boldsymbol{X}\right)$, the global reward function $R^t$ can be defined as:
\begin{equation}
	\begin{aligned}
		R^t\left(\boldsymbol{X}\right)= \sum\limits_{{n}\text{=}{1}}^{N}{r_n^t}\left(\boldsymbol{X}\right).
		%\text{=}{{P}_{n,f}^{t}}{{e}^{-\lambda\sum\limits_{{f}\text{=}{1}}^{F}({d}_{n,f}^{t}\left(\boldsymbol{X}\right)-{{Z}_{1,n,u,f}^{t}})}}
	\end{aligned}
\end{equation}
The maximum reward function in (9) only indicates the minimum local transmission delay in single agent. 
To further optimize the caching strategy, we employ the global reward function in (13) instead of the local reward function in (9).

Next, we will use the joint learning of all agents to find the globally optimal caching strategy $\boldsymbol{X^*}$. Every agent updates its target Q-values according to the observation from communication procedure. Then, every agent and its associated agents jointly update their DDQNs by sampling from experience replies.
%To find the optimal caching strategy $\boldsymbol{X}^*$, all agents learn from the environment and the communication procedure.

\subsubsection{Communication Procedure}
As the global reward function in (13) depends on the caching strategies of all agents, every agent should observe the historical caching strategies of its associated agents to adjust its own caching strategy. Thus, MARL introduces a communication procedure among agents.  
Each agent $n \in \mathcal{N}$ caches files in accordance with the current caching strategy of its associated agent $m \in \mathcal{N}_n$. We assume that agent $n$ treats the relative observation of its associated agent $m$ as the indicator of agent $n$'s caching strategy.
Let $C_{n,f}^t$ denote the number of times that the requested file $f$ has been cached in agent $n$ until time slot $t$. Agent $n$ records $C_{n,f}^t$ according to its chosen action $a_n^t$.
Then, we have $C_{-n,f}^t = \mathbb{E}_{m}[\sum\limits_{{m}\in\mathcal{N}_{n}}C_{m,f}^t]/ {t} $. 
In the communication procedure, agent $n$ collects the relative observation $C_{-n,f}^t$  and stores in the experience reply $\mathcal{D}_n$ for updating its DDQN.
\subsubsection{Update Target Q-values} 
%If file $f$ is requested during time slot $t$, the target Q-value is updated.
When file $f$ is requested in agent $n$, agent $n$ observes the historical caching strategies of its associated agents and updates its own DDQN. 
For maximizing the global reward function, we rewrite the target Q-value in (11) as follows:
\begin{equation}
	\begin{aligned}
		{\hat{Q}_n^t(\boldsymbol{s}_{n}^{t},{a}_{n}^{t}|\hat{\theta}_n)} \text{=}\dfrac{1}{C_{-n,f}^t+1}(R^t\left(\boldsymbol{X}\right)+\gamma \hat{Q}_n^t(\boldsymbol{s}_{n}^{t}, a'|\hat{\theta}_n)),
	\end{aligned}
\end{equation}
where $C_{-n,f}^t$ is an observation from which agent ${n}$ observes the historical caching strategies of its associated agents during time slot $t$.

\subsubsection{Joint Learning}
Every agent and its associated agents jointly update their own DDQNs. Single agent $n$ chooses the optimal action and stores the experience data $\left[\boldsymbol{s}_{n}^{t},{a}_{n}^{t},R^t,\boldsymbol{s}_{n}^{t+1},C_{-n,f}^t\right]$ in reply memory $\mathcal{D}_n$. Based on MARL, agent $n$ and its associated agents select randomly small batches of data from their own reply memories for updating their own DDQNs.

During each time slot, every agent learns from the interactions with environment and observes the historical caching strategies of its associated agents to choose the optimal action. After the joint learning, we can collect the joint caching space to obtain the globally optimal caching strategy. The detail of the proposed MARL based cooperative caching scheme is presented in Algorithm 1.

%reaches the new state $\boldsymbol{s}_{n}^{t+1}$ after performing the action ${a}_{n}^{t}$ under the current state $\boldsymbol{s}_{n}^{t}$ in time slot $t$ and .
%The target Q-values of agent $n$ depends on not only its own caching strategy but also the historical number of times of requested file stored in associated F-AP ${\mathcal{N}_{n}}$. The observation is defined as follows. Every agent has kept counts to record the status of cached file. When the agent $n$ caches the file $f$, $C_{n,f}^t=C_{n,f}^t+1$.
%The global reward functionThe globally optimal strategy $\boldsymbol{X}$ in problem (8) can be solved by maximizing the global reward function

%According to the above description, problem (8) is equivalent to maximize the global reward function subject to the infinite and integer constraints. Given by (9) and (13), the optimization problem in (8) can be rewritten as follows:  
%\begin{equation}
%	\begin{aligned}
%		&\max_{{x}_{n,f}}\quad {\dfrac{1}{T}}\sum\limits_{t=1}^T	R^t\left(\boldsymbol{X}\right)\\ 												       
%		&={\dfrac{1}{T}}\sum\limits_{t=1}^T\sum\limits_{{n}\text{=}{1}}^{N}\sum\limits_{{f}\text{=}{1}}^{F}{{P}_{n,f}^{t}}{{e}^{-\lambda({d}_{n,f}^{t}\left(\boldsymbol{X}\right)-{{Z}_{1,n,u,f}^{t}})}}\\
%		&\rm s.t.\quad  \left\{\begin{array}{lc}
%			\sum\limits_{f=1}^F {{x}_{n,f}} \leq {S},  \forall n \in \mathcal{N}, &\left(15\rm a\right) \\
%			{{x}_{n,f}}\in \left\{0,1\right\},  \forall n \in \mathcal{N}, \forall f \in \mathcal{F}, &\left(15\rm b\right)  \\
%			{{0<\lambda\leq1}}. &\left(15\rm c\right) \\
%		\end{array}\right.
%	\end {aligned}	
%\end{equation}

\begin{algorithm}[!t]
	\label{alg:1}
	\begin{algorithmic}[1]
		\caption{The MARL based cooperative caching scheme}
		\begin{spacing}{0.8}
			\item Initialize the reply memories $\mathsf{\mathcal{D}}_1,...,\mathsf{\mathcal{D}}_N$;
			\item Initialize the current Q-network $Q$ with the weight $\theta $, and the target Q-network $\hat{Q}$ with the weight ${\hat{\theta}}=\theta $;
			\item Initialize the count $C_{n,f}=0, n \in \mathcal{N}, f \in \mathcal{F}$;
			\For {time slot $t=1,2,...,T$}
			\For {F-AP ${n}=1,2,...,{N}$}
			\State  Collect the requested file $f$ from users in ${\mathcal{U}_n}$ ;
			\State Observe  the state $\boldsymbol{s}_{n}^{t}=\left\{{q}_{n,1}^{t},{q}_{n,2}^{t},...,{q}_{n,S}^{t},f\right\}$;
			\State Choose an action ${a_n^t}=\rm {argmax}_a Q_n^t(s,a)$ using \hspace*{2.0\dimexpr\algorithmicindent}the $\epsilon$-greedy policy under the current state  $\boldsymbol{s}_{n}^{t}$;
			\If {action ${a_n^t\ne 0}$}
			\State Update ${C_{n,f}=C_{n,f}+1}$;
			\State Update ${C_{n,{a_n^t}}=0}$;
			\State Execute the action ${a_n^t}$, and replace the \hspace*{3.0\dimexpr\algorithmicindent}${a_n^t}$-th stored file in F-AP $n$ with file $f$;
			%\State $\boldsymbol{s}_{n}^{t+1}=\left\{{q}_{n,1}^{t},...,{q}_{n,a_n^t}^{t}=f,...,{q}_{n,S}^{t},f\right\}$;
			\EndIf
			\State Compute ${C_{-n,f}^t}=\mathbb{E}_{m}[\sum\limits_{{m}\in\mathcal{N}_{n}}C_{m,f}^t]/t $; 
			\State Save $\left[\boldsymbol{s}_{n}^{t},{a}_{n}^{t},R^t,\boldsymbol{s}_{n}^{t+1},C_{-n,f}^t\right]$ in ${\mathcal{D}_{n}}$;
			\While {F-AP $m \in \mathcal{N}_{n} \cup \left\{n\right\}$}
			\State Obtain the reward ${R^t}\left(\boldsymbol{X}\right)$ according to (13);
			\State Randomly sample a mini-batch of experiences \hspace*{3.0\dimexpr\algorithmicindent}from ${\mathcal{D}_m}$;
			\State Update the target Q-values ${\hat{Q}_m^t(\boldsymbol{s}_{m}^{t},{a}_{m}^{t}|\hat{\theta}_m)}$ \hspace*{3.0\dimexpr\algorithmicindent}according to (14);
			\State Update the weight ${\theta_m}$ by the loss function \hspace*{3.0\dimexpr\algorithmicindent}$L(\theta_m)$ according to (12);
			\State Reset ${\hat{\theta}_{m}}={\theta}_{m} $ every $\nu$ time slots;
			\EndWhile
			%\EndIf
			%\EndIf	
			\EndFor
			\State Obtain the caching strategy $\boldsymbol{X^*}$ according to the joint \hspace*{1.0\dimexpr\algorithmicindent}state space ${\mathcal{S}={S_1\times ... \times S_N}}$;
			\EndFor
		\end{spacing}
	\end{algorithmic}
	\vspace{-1.0em}
\end{algorithm}

\section{Simulation Results}

The performance of the proposed MARL-based cooperative caching scheme is evaluated via simulations. The users' file preference follows the Mandelbrot-Zipf distribution with the skewness factor ${\tau_t = 1.1}$. The small-scale channel gain ${|{h}_{a,b}^{t}|^2}$ follows standard exponential distribution. The bandwidth $B$ is set to 100MHz\cite{CachePlacement}. Each F-AP serves the users in a circular cell with a radius of 100m. Assume that no inter-cell interference is induced. The file size is set to 1Mbits. For simplification, the transmission rate in backhaul link is set to $R=$100Mbps. The learning rate $\alpha$ is set to 0.001 and the reward decay $\gamma$ is set to 0.9. Unless otherwise stated, we set $U=50, F=500, N=5$. In the simulations, the traditional scheme (Least Recently Used (LRU)) and the learning schemes (DQN and Independent Q-learning (IQL)) are chosen as the benchmark schemes.

\begin{figure}[t]
	\centering
	\includegraphics[height=5.0cm,width=8.0cm]{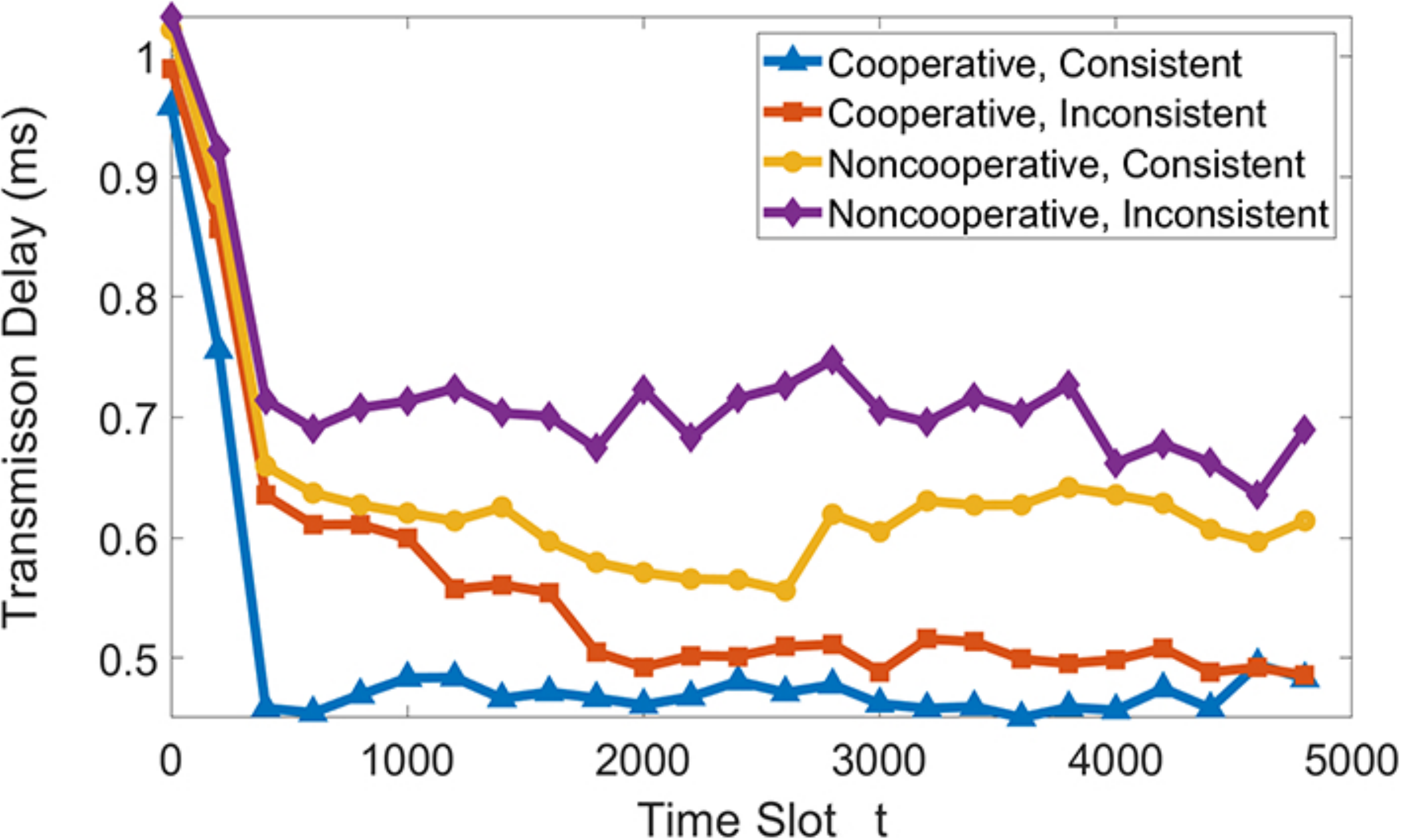}\\
	\caption{Transmission delay versus different caching and different user preference.}
	\label{1}
\end{figure}
In Fig. 3, we show the delay performance of different caching and different user preference$^1$
\footnotetext[1]{For consistent user preference, we set the
random permutation $\mathcal{\Phi}_u^t$ as a constant. And for inconsistent user preference, we set the random permutation $\mathcal{\Phi}_u^t$ as a time-varying random permutation of $\mathcal{F}$.}based on MARL. It can be observed that the four schemes can approach their stable transmission delay as time slot increases. The noncooperative caching schemes have higher transmission delay than the cooperative caching schemes. The reason is that F-APs need to fetch more files from the cloud server in noncooperative caching schemes. It can also be observed that the transmission delay has the lowest value in the cooperative caching and consistent user preference scheme. That is because our proposed scheme can learn the user preference and get the content popularity of every F-AP.
\begin{figure}[t]
	\centering
	\includegraphics[height=5.5cm,width=7.5cm]{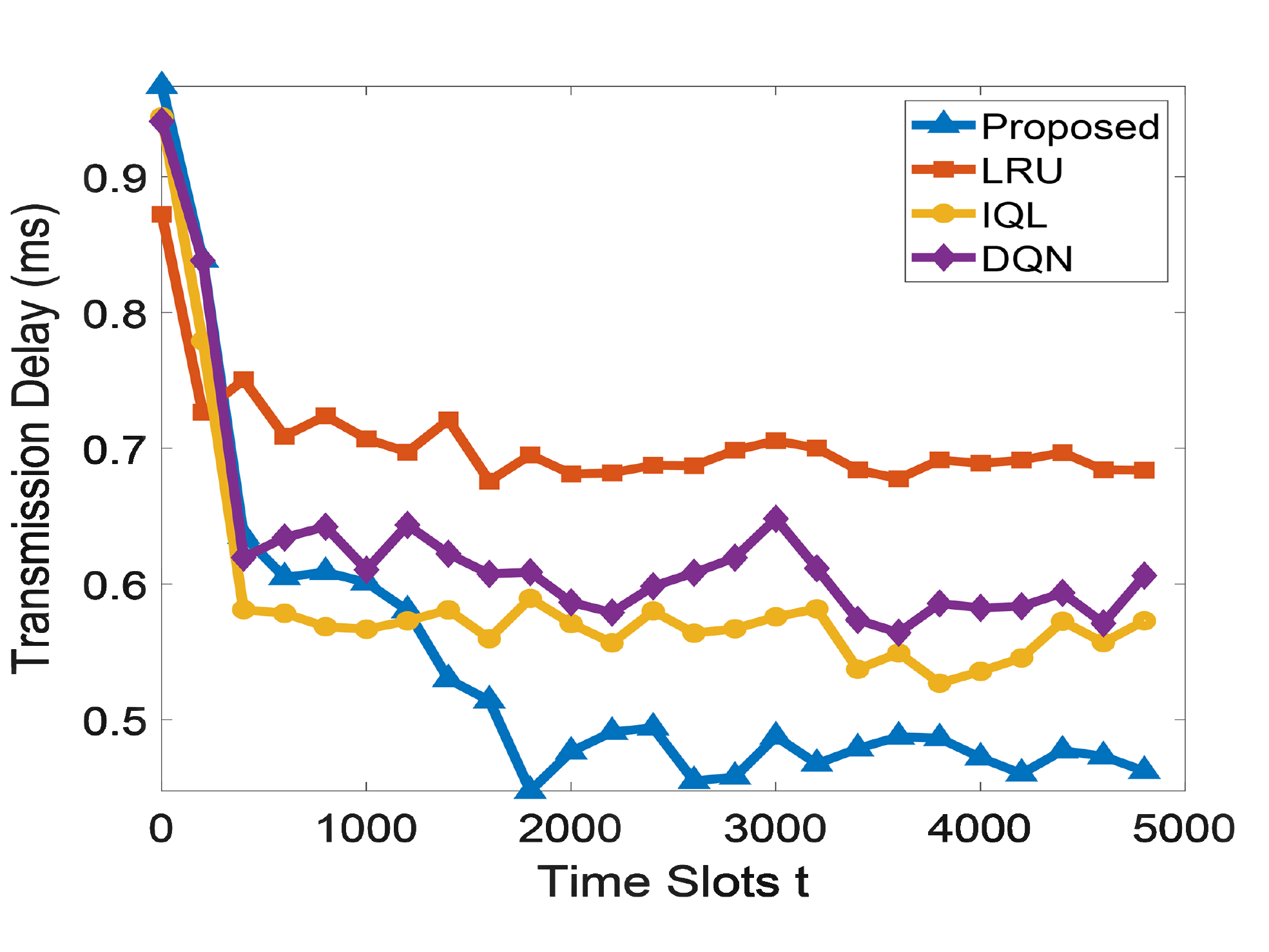}\\
	\caption{Transmission delay versus time slot for the proposed scheme and three benchmark schemes.}
	\label{1}
\end{figure}

In Fig. 4, we show the convergence performance of our proposed scheme in comparison with the three benchmark schemes. It can be observed that our proposed scheme converges to a relatively stable value when time slot $t$ is larger than 2000. Compared with the benchmark schemes, our proposed scheme has lower convergence speed and better delay performance. 
The reason is that our proposed scheme has few records about the historical caching strategies at the beginning of the training. 
%That reason is that our proposed 
With the continuous caching updates, our proposed scheme can gradually leverage the cooperation among F-APs and find the globally optimal caching scheme. Meanwhile, LRU has the highest transmission delay as no learning is adopted. IQL and DQN have the close delay performances since they neglect the interactions among agents.

\begin{figure}[t]
	\centering
	\includegraphics[height=5.5cm,width=7.5cm]{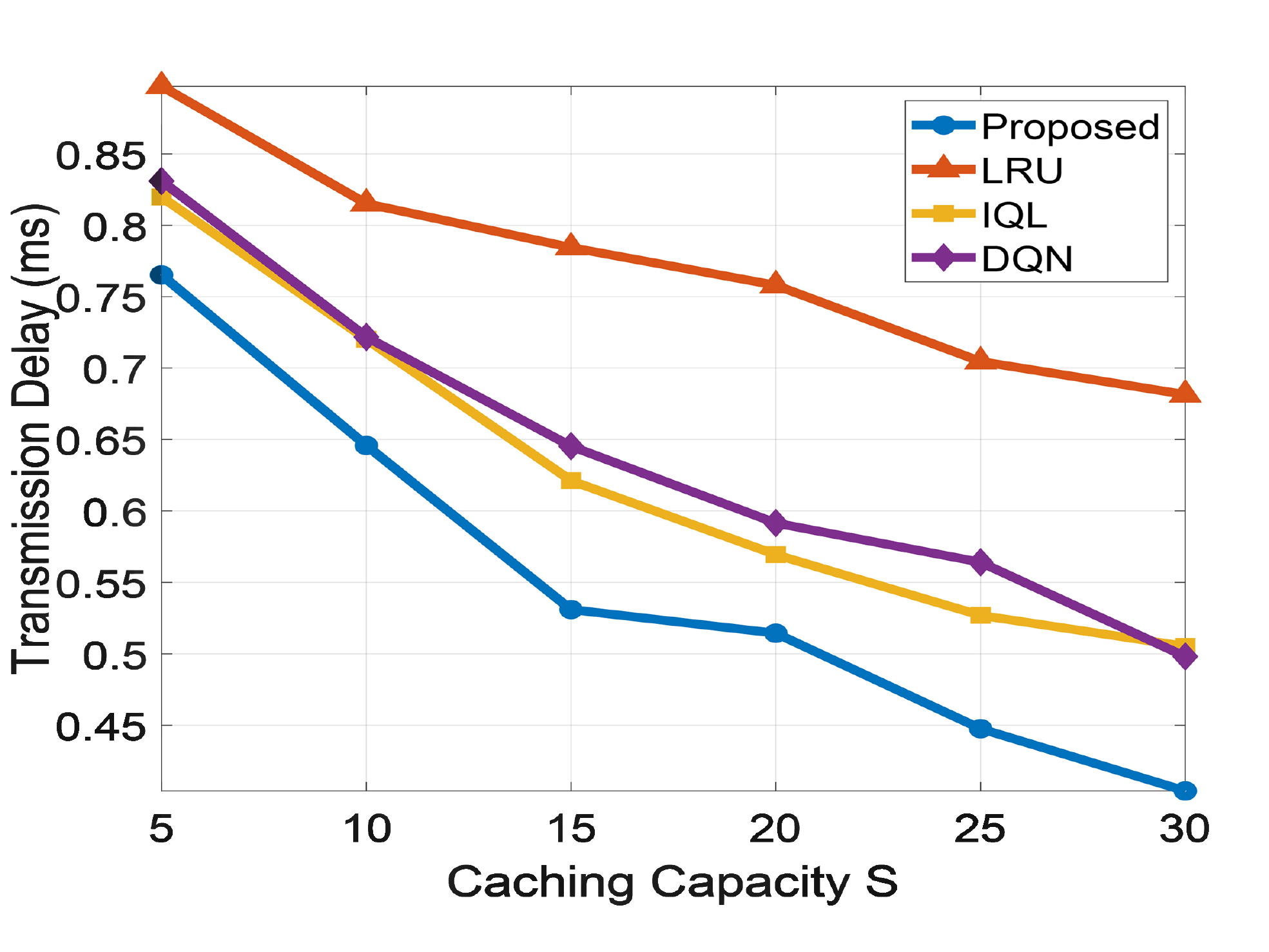}\\
	\caption{Transmission delay versus cache capacity for the proposed scheme and three benchmark schemes.}
	\label{1}
\end{figure}

In Fig. 5, we show the transmission delay of our proposed scheme and the benchmark schemes while varying the F-AP caching capacity. It can be observed that the transmission delay reduces as the caching capacity increases. It can also be observed that the transmission delay of our proposed scheme is always lower than that of the benchmark schemes. That is reasonable because larger caching capacity enables F-APs to cache more popular files simultaneously and our proposed scheme can utilize the communication among F-APs to reduce the average transmission delay.

\section{Conclusions}

%In this paper, we have studied the cooperative edge caching scenario in F-RANs. We have formulated the corresponding optimization problem to minimize average transmission delay of the entire system.
%In each F-AP, DDQN is applied to update its own caching contents.
%By introducing the communications among F-APs,  we have proposed an MARL-based caching strategy to find the globally optimal caching strategy.   
In this paper, we have proposed an MARL-based cooperative caching scheme in F-RANs. %We formulate the corresponding optimization problem to minimize the average transmission delay of the entire system.
In each F-AP, the DDQN has been utilized to meet the integer and cache capacity constraints.
%MARL can employ the DDQNs in every F-AP and utilize the communication procedure for the global optimal caching strategy. 
%MARL can enhance the cooperation among F-APs and improve the caching performance by introducing the communication process. 
%MARL can enhance the cooperation among F-APs and improve the caching performance by introducing the communication process.
In addition, MARL has introduced the communication procedure to leverage the cooperation among F-APs. 
By recording the historical strategies of the associated F-APs, our proposed scheme has made agents communicate with other agents to maximize the global reward function and reduce the average transmission delay further.
%Our proposed strategy has reduced the average transmission delay of the entire system under the joint learning of all agents.
Simulation results have shown that our proposed scheme achieves a significant performance improvement compared with the benchmark schemes. %in reducing the average transmission delay.

%For finding the  optimal caching strategy, the local transmission delay in single F-AP is regarded as an MDP. We utilize DDQN to make locally optimal caching strategies. In addition, a communication process among F-APs is applied to take full account of the caching state of associated F-APs. Through the iterative exchange of information, our proposed strategy can find the globally optimal caching strategy and the global transmission delay can be reduced. Our caching strategy has considerably utilized cooperation among F-APs and significantly reduces the average transmission delay in F-RANs.  
%In this paper, we study the cooperative caching problem for F-RANs. In the Through the iterative communications among F-APs, the average transmission delay can be reduced and global optimization problem is tackled dynamically.

\section*{Acknowledgements}
This work was supported in part by the Natural Science Foundation of China under grant 61971129, the Natural Science Foundation of Jiangsu Province under grant BK20181264, the Shenzhen Science and Technology Program under Grant KQTD20190929172545139 and JCYJ20180306171815699, and the National Major Research and Development Program of China under Grant 2020YFB1805005.

\balance
\bibliographystyle{IEEEtran}
\bibliography{reference}

% Generated by IEEEtran.bst, version: 1.14 (2015/08/26)
\begin{thebibliography}{10}
\providecommand{\url}[1]{#1}
\csname url@samestyle\endcsname
\providecommand{\newblock}{\relax}
\providecommand{\bibinfo}[2]{#2}
\providecommand{\BIBentrySTDinterwordspacing}{\spaceskip=0pt\relax}
\providecommand{\BIBentryALTinterwordstretchfactor}{4}
\providecommand{\BIBentryALTinterwordspacing}{\spaceskip=\fontdimen2\font plus
\BIBentryALTinterwordstretchfactor\fontdimen3\font minus
  \fontdimen4\font\relax}
\providecommand{\BIBforeignlanguage}[2]{{%
\expandafter\ifx\csname l@#1\endcsname\relax
\typeout{** WARNING: IEEEtran.bst: No hyphenation pattern has been}%
\typeout{** loaded for the language `#1'. Using the pattern for}%
\typeout{** the default language instead.}%
\else
\language=\csname l@#1\endcsname
\fi
#2}}
\providecommand{\BIBdecl}{\relax}
\BIBdecl

\bibitem{AComprehensiveSurvey}
M.~A. Habibi, M.~Nasimi, B.~Han, and H.~D. Schotten, ``A comprehensive survey
  of {RAN} architectures toward {5G} mobile communication system,'' \emph{IEEE
  Access}, vol.~7, pp. 70\,371--70\,421, May 2019.

\bibitem{SocialAware}
X.~Wang, S.~Leng, and K.~Yang, ``Social-aware edge caching in fog radio access
  networks,'' \emph{IEEE Access}, vol.~5, pp. 8492--8501, Apr. 2017.

\bibitem{Fog-computing-based}
M.~Peng, S.~Yan, K.~Zhang, and C.~Wang, ``Fog-computing-based radio access
  networks: {Issues} and challenges,'' \emph{IEEE Network}, vol.~30, no.~4, pp.
  46--53, Jul. 2016.

\bibitem{APigeonInspired}
C.~Xia, Y.~Jiang, M.~Peng, F.-C. Zheng, M.~Bennis, and X.~You, ``Cooperative
  edge caching in fog radio access networks: A pigeon inspired optimization
  approach,'' in \emph{2019 IEEE Global Communications Conference (GLOBECOM)},
  Feb. 2019, pp. 1--6.

\bibitem{BrainStorm}
Y.~Jiang, X.~Chen, F.-C. Zheng, D.~Niyato, and X.~You, ``Brain storm
  optimization-based edge caching in fog radio access networks,'' \emph{IEEE
  Transactions on Vehicular Technology}, vol.~70, no.~2, pp. 1807--1820, Jan.
  2021.

\bibitem{Q-learning}
C.~Wang, S.~Wang, D.~Li, X.~Wang, X.~Li, and V.~C.~M. Leung, ``Q-learning based
  edge caching optimization for {D2D} enabled hierarchical wireless networks,''
  in \emph{2018 IEEE 15th International Conference on Mobile Ad Hoc and Sensor
  Systems (MASS)}, Oct. 2018, pp. 55--63.

\bibitem{LearningAutomata}
Z.~Yang, Y.~Liu, Y.~Chen, and L.~Jiao, ``Learning automata based {Q-Learning}
  for content placement in cooperative caching,'' \emph{IEEE Transactions on
  Communications}, vol.~68, no.~6, pp. 3667--3680, Mar. 2020.

\bibitem{Dueling}
B.~Guo, X.~Zhang, Q.~Sheng, and H.~Yang, ``Dueling deep-{Q}-network based
  delay-aware cache update policy for mobile users in fog radio access
  networks,'' \emph{IEEE Access}, vol.~8, pp. 7131--7141, Jan. 2020.

\bibitem{DistributedEdge}
J.~Yan, Y.~Jiang, F.~Zheng, F.~R. Yu, X.~Gao, and X.~You, ``Distributed edge
  caching with content recommendation in fog-rans via deep reinforcement
  learning,'' in \emph{2020 IEEE International Conference on Communications
  Workshops (ICC Workshops)}, Jul. 2020, pp. 1--6.

\bibitem{DeepReinforcement}
L.~Li, Y.~Xu, J.~Yin, W.~Liang, X.~Li, W.~Chen, and Z.~Han, ``Deep
  reinforcement learning approaches for content caching in cache-enabled {D2D}
  networks,'' \emph{IEEE Internet of Things Journal}, vol.~7, no.~1, pp.
  544--557, Nov. 2020.

\bibitem{CooperativeEdgeCaching}
M.~Zhang, Y.~Jiang, F.-C. Zheng, M.~Bennis, and X.~You, ``Cooperative edge
  caching via federated deep reinforcement learning in fog-rans,'' in
  \emph{2021 IEEE International Conference on Communications Workshops (ICC
  Workshops)}, Jul. 2021, pp. 1--6.

\bibitem{Multi-Agent}
K.~Jiang, H.~Zhou, D.~Zeng, and J.~Wu, ``Multi-agent reinforcement learning for
  cooperative edge caching in internet of vehicles,'' in \emph{2020 IEEE 17th
  International Conference on Mobile Ad Hoc and Sensor Systems (MASS)}, Dec.
  2020, pp. 455--463.

\bibitem{CachePlacement}
J.~Liu, B.~Bai, J.~Zhang, and K.~B. Letaief, ``Cache placement in fog-rans:
  From centralized to distributed algorithms,'' \emph{IEEE Transactions on
  Wireless Communications}, vol.~16, no.~11, pp. 7039--7051, Aug. 2017.

\bibitem{Citations}
Z.~Silagadze, ``Citations and the {Zipf}-{Mandelbrot} law,'' \emph{COMPLEX
  SYSTEMS -CHAMPAIGN-}, vol.~11, no.~6, pp. 487--500, Sep. 1997.

\bibitem{AMeanField}
Y.~Jiang, Y.~Hu, M.~Bennis, F.-C. Zheng, and X.~You, ``A mean field game-based
  distributed edge caching in fog radio access networks,'' \emph{IEEE
  Transactions on Communications}, vol.~68, no.~3, pp. 1567--1580, Dec. 2020.

\bibitem{DoubleQlearning}
H.~{van Hasselt}, A.~{Guez}, and D.~{Silver}, ``{Deep Reinforcement Learning
  with Double Q-learning},'' \emph{arXiv e-prints}, p. arXiv:1509.06461, Sep.
  2015.

\end{thebibliography}

\end{document}